\documentclass[pss]{wiley2sp} 
\usepackage{amsmath}
\usepackage{verbatim}

\begin{document}

\title{Kondo effect in oscillating molecules}

\titlerunning{Kondo effect in oscillating molecules}

\author{%
  Jernej Mravlje\textsuperscript{\textsf{\bfseries 1\Ast}},
  Anton Ram\v{s}ak\textsuperscript{\textsf{\bfseries 2,1}}
}
\authorrunning{J. Mravlje et al.}

\mail{e-mail
  \textsf{jernej.mravlje@ijs.si}}

\institute{%
  \textsuperscript{1}\,Jo\v zef Stefan Institute, Jamova 39, 1000
  Ljubljana, Slovenija\\
  \textsuperscript{2}\,Faculty of Mathematics and Physics, University
  of Ljubljana, Jadranska 19, 1000 Ljubljana, Slovenija\\
}

\received{XXXX, revised XXXX, accepted XXXX} 
\published{XXXX} 


\pacs{72.15.Qm,73.23.-b,73.22.-f}

\abstract{%
%
%
%
\abstcol{%
  We consider electronic transport through break-junctions bridged by
  a single molecule in the Kondo regime. We describe the system by a
  two-channel Anderson model. We take the tunneling matrix 
  elements to depend on the position of the molecule. It is shown, that if
  the modulation 
  of the tunneling by displacement is large, the
  potential confining the molecule to the central position between the
  leads is softened and the
  position of the  
  molecule is increasingly susceptible to external
  perturbations that break the inversion symmetry. In this regime, the
  molecule is attracted to one of the leads and as a consequence the
  conductance is small.  We argue on semi-classical grounds why the
  softening  occurs and
  corroborate our findings by numerical examples obtained by Wilson's
  numerical renormalization group and Sch\"onhammer-Gunnarsson's
  variational method.}{%
  }
}

%
%
\titlefigure[height=3.1cm,keepaspectratio]{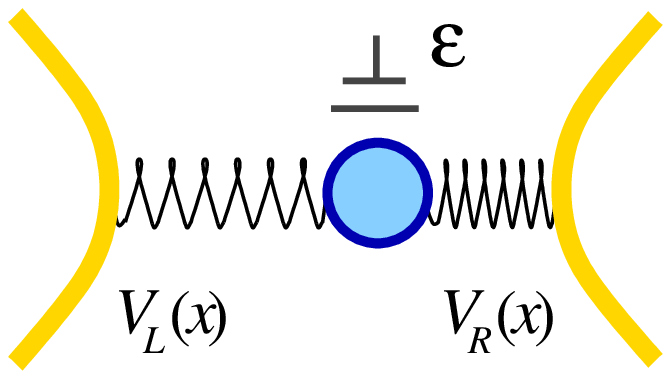}
\titlefigurecaption{%
Sketch of the model device. The overlap integrals between the
molecular orbital and the leads ($V_L,V_R$ for left, right lead,
respectively) are modulated by the position of the
molecule. The energy of the molecular level $\epsilon$ is determined
by the gate voltage. }

\maketitle   

\section{Introduction and model}
In recent years considerable advance has been achieved in manufacturing
mesoscopic systems, which due to their tunability with external
electrodes provide a playground for research in the
correlated electron systems. For example, the Kondo effect -- generic
name for phenomena related to increased scattering off impurities with
internal degrees of freedom --  was
observed in measurements of electron transport through quantum
dots \cite{goldhaber98}, atoms, and 
molecules
\cite{madhavan98,liang02,park02,yu04_2,pasupathy05,zhao05,yu05}.

Unlike in experiments with quantum dots, the transport through break
junctions is strongly affected by the molecular vibrational modes,
because the frequency of the oscillations is of comparable magnitude
than other energy scales, such as Coulomb repulsion.
 For example,
 the side-peaks in the 
non-linear conductance \cite{park02,yu04_2,pasupathy05} were observed
indicating the transfer of energy from the oscillations to the electron
current. By comparing the observed frequencies to the frequencies of
the molecular internal modes it was shown \cite{park02},
that in some cases also the oscillations of the molecule with respect to the
leads have to be taken into account.

In this work we concentrate on such a case. The tunneling
is generally dependent on the overlap between the
wave-functions.  This motivates us to investigate the effects of
the modulation of the tunneling  matrix elements by molecular
oscillations. 

More specifically,
we describe the break junction by  a model consisting of two
metallic leads (half-filled bands of non-interacting electrons). The
leads are  
bridged by a single molecule which we assume is confined
harmonically to the center between the leads. The position of the
molecule $x$  determines the tunneling
matrix elements. Assuming the leads are identical and the displacement
of the molecule from the center of inversion $x$ is
small \cite{mravlje08}, the tunneling matrix
elements towards left and right leads read
$V_{L,R}(x)=(1\mp g x)V$. 

The Hamiltonian consists of several parts
\begin{equation} \label{eq:hami} H=H_L +H_R +  H_\mathrm{mol}+ H_{\mathrm{vib}} +
H'. \end{equation} 
The first two terms describe the isolated leads
\begin{equation} H_{\alpha} = \sum_{k \alpha\sigma} \epsilon_k
c_{k\alpha \sigma} ^\dagger c_{k\alpha \sigma},
\end{equation} $\alpha=L,R$ is the
lead index, $\sigma=\uparrow, \downarrow$ is the spin index and $k$ the
wave-vector index; $c^{\dagger}$ (as well as $d^{\dagger}$ to be 
introduced below) denote the fermion creation operators. The precise
form of the band dispersion $\epsilon_k$ is not important for the results presented here,
provided the density of states is smooth and symmetric around the Fermi
surface; nevertheless for definiteness we note that in our
calculations we used flat (NRG simulations) and
tight-binding bands (simulations with Sch\"onhammer-Gunnarsson wave functions).

The second
term  
describes the molecular orbital (we assume a single molecular level
participates actively in the transport),
\begin{equation} 
H_{\mathrm{mol}}= \epsilon (n_\uparrow + n_\downarrow) + U n_\uparrow n_\downarrow, 
\end{equation}
where $n_\sigma = d^\dagger_\sigma d_\sigma$ counts the number of electrons occupying
the orbital, $\epsilon$ the energy of the orbital relative to the
chemical potential of the leads, $U$ is the Coulomb
repulsion. We concentrate on the particle-hole symmetric case
$\epsilon=-U/2$. $H_\mathrm{vib}=\Omega a^\dagger a$  describes the phonon
mode ($a^\dagger$ is the phonon creation operator). 

The coupling between the leads, the
molecular orbital and also phonons is described by
\begin{eqnarray}
H'= V\sum_{k\sigma} & & \left[    (1-\zeta-gx)  c^\dagger_{kL \sigma}d_\sigma+
\right. \\
   & & {} \left.  +\; 
   (1+\zeta + gx) c^\dagger_{kR \sigma} d_\sigma \right]
   + h.c.  \nonumber
\end{eqnarray}
Here we introduced a constant $\zeta$, which we will use to test for the
influence of the breaking of the inversion symmetry; for $\zeta=0$ the
Hamiltonian is symmetric with respect to operation $x \to -x,
L\leftrightarrow R$; finite $\zeta$ breaks this symmetry. 

In the model with no coupling to phonons ($g=0$) for $\zeta=0$ only the even
combination of the operators, {\it i.e.} $c_e = (c_L + c_R)/\sqrt{2}$
(other indeces are suppressed) in the leads is
coupled to the molecular 
orbital. The life time of  electrons on the orbital is finite due to
the tunneling to the leads; the
hybridization $\Gamma$ (inverse life-time)  for the flat band reads
$\Gamma = 2\pi 
\rho V^2$, where $\rho$  is the density of
states in each of the two leads (for flat band of half-width $D$,
$\rho=1/(2D)$. The odd linear combinations $c_o=( 
c_L - c_R)/\sqrt{2}$ are decoupled. For $g=0$ and $\zeta>0$ still a
particular linear combination of the states in the leads is decoupled
and the system can be described by the single-channel model. 

For finite $g\neq 0$ this no longer holds and the molecule is  coupled
to both 
conduction channels.  Rewriting the coupling term in the even-odd
basis, 
\begin{equation}
H' =  \sqrt{2}V\sum_{k\sigma} \left[ c^\dagger_{ke \sigma}d_\sigma  +
 (\zeta + g x) c^\dagger_{ko \sigma} d_\sigma\right] + h.c., 
\end{equation}
makes it manifest that we are dealing with a two-channel
Anderson model.
No linear combination of the conduction
electrons can be integrated out because the coupling to the odd channel is
mediated by phonons. This occurs because  the modulation of tunneling
is antisymmetric with respect to inversion; if the modulation of
tunneling is symmetric \cite{cornaglia05a,mravlje05} the orbital is
still coupled only to the even channel.

It is difficult to access the low temperature (Kondo) regime   because
of the presence of  
exponentially small energy scale $T_K \propto \exp \left[-1/ (\rho
  J\right)]$ ($J\sim V^2/U$ 
is the magnitude of the exchange coupling). Only few methods \cite{bonca04} 
reproduce the increase of conductance towards
the unitary limit $G_0=2e^2/h$ for temperatures smaller than $T_K$
accurately.   
On the other hand, in the high-temperature regime (relevant for  
nanoelectromechanical systems \cite{novotny04,twamley06,johansson08})
it is adequate to ignore the Kondo correlations and take only
lowest orders in tunneling into account. 

For the work on this model in the
Kondo regime, which was originally stimulated by Ref. \cite{alhassanieh05},
we refer the reader to
Refs. \cite{balseiro06,mravlje06,mravlje07,mravlje08}. Another very
recent paper discusses a similar model as an example of
two-level system \cite{lucignano08}.

In the following section we first demonstrate that the confining
potential is weakened by the electron-phonon coupling and can even be 
driven to a form of the double-well. Then we give numerical examples
on how the emergence of the double-well potential affects static
and dynamic properties of the molecule. Finally, we show that when the
inversion symmetry is not perfect ($\zeta\neq 0$), the 
electron-phonon coupling will drive the molecule away from the central
position. As a consequence, in this regime the conductance can be
significantly suppressed. 

\section{The emergence of a double well potential}
\begin{figure}[t]%
\includegraphics*[width=0.8 \linewidth,keepaspectratio]{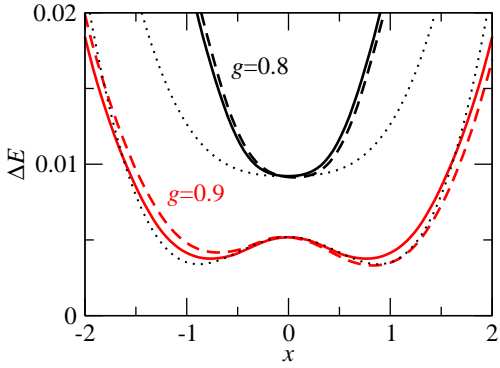}
\caption{%
Numerical results for effective potential: $\zeta=0$ (full lines),
$\zeta=0.01$ (dashed lines). Semi-classical estimate is also shown 
(dotted). Here and in subsequent figures we use $U=0.3$,
$\Gamma=0.02$, $\Omega=0.2$. We use the band half-width 
$D$ as the energy unit. The curves are shifted vertically so that the
values at $x=0$ match.
\label{fig2}
}
\end{figure}
It is easy
to demonstrate that a double well effective oscillator potential may
form under the influence of a sufficient electron-phonon coupling
$g>g_\mathrm{d}$, where we define the delimiting value of the coupling
constant. For $U=0$ we can make a simple estimate on the form of the
effective oscillator potential by a substitution $a, \sim a^\dagger
\to x/2$, where $x$ is a real-valued constant. In the wide-band limit
($\Gamma/D$ small) the energy gain due to the hybridization
is \cite{fabrizio07} $\Delta E_{\mathrm{hyb}}=-2/\pi \tilde{\Gamma}
\log D/\tilde{\Gamma}$, where we use the effective
displacement-dependent hybridization $\tilde{\Gamma}(x)=\Gamma
(1+g^2x^2)$. The elastic energy cost is $\Delta E_{\mathrm{el}}=\Omega
x^2/4$, hence in this semi-classical approximation the dependence of
energy $\Delta E_\mathrm{SC}=\Delta E_{\mathrm{el}}+ \Delta
E_{\mathrm{hyb}}$ on $x$ can be written in a closed form
\begin{equation} \Delta
E_\mathrm{SC}(x)=\Omega x^2/4 -(2/\pi) \tilde{\Gamma}(x)  \log
  \{D/[\tilde \Gamma (x)]\}.
  \label{eq:enerSC}
\end{equation}
 The prefactor of the $x^2$-term in the small-$x$ expansion is equal
 to $\Omega/4 - \left\{(2/\pi) g^2 \Gamma \left[\log(D/\Gamma)
   -1\right]\right\}$, hence for increasing $g$ the elastic potential
 is softened and a double well effective potential emerges for 
\begin{equation}
    g>g_\mathrm{d} = \sqrt{\frac{\pi\Omega}{8 \Gamma
    \left[\log(D/\Gamma)-1\right]}}.
\end{equation}
 We plot $\Delta E_\mathrm{SC} (x)$ for
 $g=0.8<g_\mathrm{d}$ and $g=0.9>g_\mathrm{d}$ in Fig.~\ref{fig2}
 (thin, dotted lines). 

 We estimate the effective potential also numerically using a
variational method based on the Sch\"on\-hammer-Gunnar\-sson  \cite{schonhammer76,gunnarsson85}
wave fuction (the details of our implementation
are given in our previous work
\cite{rejec03a,rejec03b,mravlje05}). Briefly, the idea is to find an auxiliary
non-interacting Hamiltonian $\tilde{H}$ [of the same  
form as $H$ in Eq.~(\ref{eq:hami}), but for $g=0, U=0$ and renormalized parameters
$\tilde{V}_L,\tilde{V}_R,\tilde{\epsilon}$], which minimizes the 
variational ground state energy $E=\langle\Psi | H|\Psi\rangle$.
The variational function $\Psi$ is expressed in the basis of
projection operators $P_{i}$ acting on the 
Hartree-Fock ground state  $|\Psi_0 \rangle$ (which includes the
phonon vacuum) of the auxiliary
Hamiltonian $\tilde{H}$,
\begin{equation} |\Psi\rangle = \sum_{ni} \psi_{ni} (a^{\dagger})^n P_i |\Psi_0
  \rangle.
\end{equation}
To obtain the ground state, we minimize energy with respect to all the
parameters of $\tilde{H}$. On the other hand, by restricting the minimization to a
particular subspace (for example, by fixing the ratio $V_L/V_R=r$) we
obtain the variational wave-function $\Psi_r$ for which the
expectation value $\langle x \rangle_r $ is a function of $r$. The
pairs $(\langle x \rangle_r, E)$ constitute our estimate of the
effective potential and are plotted in Fig.~\ref{fig2} for $\zeta=0$
(full lines) and $\zeta=0.01$ (dashed lines). The agreement between
the semi-classical estimate and numerical results is reasonable.

The perturbation $\zeta=0.01$ breaks the inversion symmetry, therefore
the right minima in $g>g_\mathrm{d}$ regime in this case is lower in
energy. In this regime, the molecule will predominantly reside near
the right lead. Conversely, for $g<g_\mathrm{d}$ the potential is only
slightly perturbed.

\section{NRG results}
Having established well the emergence of the double well effective
potential we now
check how it is reflected first in the static properties and then the 
dynamical response of the system. To obtain these quantities, we have
performed the numerical simulations using the well known Wilson's
numerical renormalization group \cite{wilson75,bulla08}(NRG)
method. We restrict ourselves to the limit of zero temperature ($T
\to 0$). The details of the calculations are given in
Ref.~\cite{mravlje08}.

\begin{figure}[t]%
\includegraphics*[width=0.8 \linewidth,keepaspectratio]{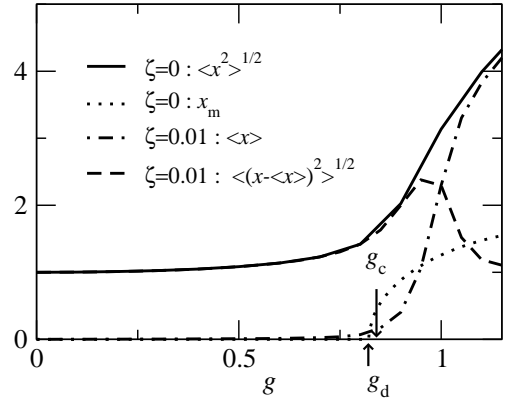}
\caption{%
Average displacement and displacement fluctuations obtained by NRG
compared to the semi-classical estimate $x_{\mathrm{m}}$ (dotted). 
\label{fig3}
}
\end{figure}

\subsection{Static quantities}
We begin by looking at the static quantities. The average displacement
$\langle x \rangle $ for $\zeta=0$ vanishes (as expected for an
operator of odd parity under inversion in a state of well-defined
parity). The fluctuations of displacement $(\Delta x^2)^{1/2}=\langle
(x-\langle x \rangle )^2\rangle^{1/2}$, shown in Fig.~\ref{fig3} (full line)
increase monotonically with $g$. The slope of $(\Delta x^2)^{1/2}$ is
increased considerably at $g\sim g_\mathrm{d}$ [or $(\sim
  g_\mathrm{c})$, see Ref.~\cite{mravlje08}],
where the double well like effective potential is formed.  This change
of slope is driven by the increased 
hybridization in the odd-channel.

For $\zeta=0.01$ the absence of inversion symmetry is reflected in the
nonvanishing average displacement (dashed-dotted), which monotonically
increases with increasing $g$. Therefore the fluctuations of
displacement (dashed) in this case reach a maximum and then decrease
with increasing $g$.

 For comparison, we plot also  the
position of the minimum of the potential $x_\mathrm{m}$ (dotted line)
obtained from the semi-classical estimate Eq.~(\ref{eq:enerSC}), 
 \begin{equation} 
    x_\mathrm{m} = \sqrt{\frac{\pi\Omega(g-g_\mathrm{d})}{4 \Gamma
   g_\mathrm{d}^5}}. 
\label{eq:semiclassx}
\end{equation}

 \subsection{Phonon propagator}
 \begin{figure}[t]%
\includegraphics*[width=0.8 \linewidth,keepaspectratio]{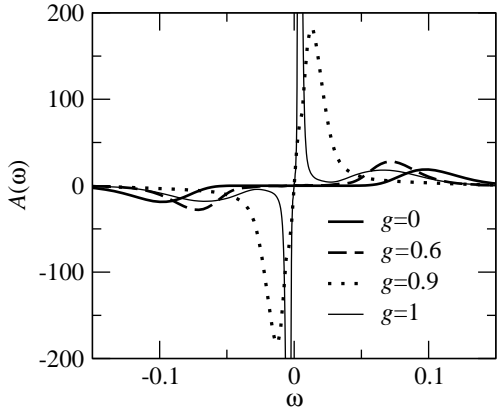}
\caption{%
Displacement spectral functions for $\zeta=0$. 
\label{fig4}}
\end{figure}

Now we turn to the renormalization of the phonon propagator by the
electron-phonon coupling. The dynamical information about oscillator
is contained in the 
displacement Green's function. The displacement spectral function 
\begin{align} \mathcal{A}(\omega)&=-\frac{1}{\pi} \mathrm{Im} \ll x,x \gg_\omega
  = \nonumber \\
&=-\frac{1}{\pi}\mathrm{Im}  
\int_{0}^{\infty} (-i) \langle[x(t),x(0)]\rangle e^{i \omega t}
dt \end{align} is an  odd function
of $\omega$ due to the hermiticity of $x$. Since $\mathcal{A}(\omega)$ is
  in NRG evaluated for a finite system it consists of several
  $\delta$-peaks of different weights. To obtain a smooth spectral
  function we have used the
  Gaussian broadening on the logarithmic scale \cite{bulla01}, where
  the Dirac $\delta$ function is broadened according to 
\begin{equation}\delta(\omega - \omega_n) \to \frac{1}{b \omega_n \pi}
\exp \left\{ -\left[\frac{\log (\omega/\omega_n)}{b}\right]^2 -\frac{b^2}{4}\right\}, \end{equation}
and we used $b=0.3$ in our calculations.

In Fig.~\ref{fig4} we plot $\mathcal{A}(\omega)$ for various $g$ and
$\zeta=0$. The width of 
the high frequency peaks is overestimated due to the
broadening procedure described above (for example, the width of the
peak at $\omega=\Omega$ for $g=0$ should vanish).    We could
use the Dyson equation \cite{bulla98,jeon03} to obtain sharper peaks
but we avoid this complication because on one hand there is no {\it a
  priori} guarantee that such a 
procedure gives more accurate results for large $g$ 
and on the other hand in this work we are interested only in the
position and not the width of the peaks.

  For intermediate $g$ (starting at $g\sim
  0.5$ for the parameters  used here) the vibrational mode
  begins to soften; the characteristic frequency of the oscillations
  is decreased. At  still larger $g> g_{\mathrm{d}}$  two peaks
  emerge. The high frequency peak corresponds to the oscillations
  within each of the minima of the double-well potential, and the
  low-frequency peak (we denote its position by
  $\omega_0$) corresponds to the slow tunneling between the
  degenerate (or near-degenerate for $\zeta>0$) minima.

The propagators  for finite $\zeta$ and $\zeta=0$ look alike, provided
$g$ is small enough that $\omega_0$ does not
decrease below the frequency given by the energy difference $\propto \zeta$
between the minima of the two wells. 

\begin{figure}[t]%
\includegraphics*[width=0.8\linewidth,keepaspectratio]{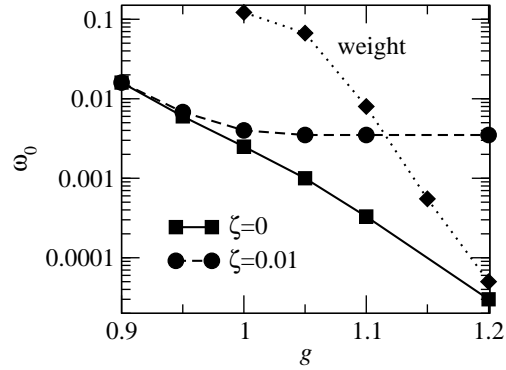}
\caption{%
  The frequency  of the soft mode
  peak  as a function of
  $g$.  The weight of the soft  mode peak for $\zeta=0.01$ and normalized to some arbitrary value (dotted). 
\label{fig5}}
\end{figure}

For larger $g$  the high frequency behaviour remains similar but
as shown in Fig.~\ref{fig5} $\omega_0$ which decreases for $\zeta=0$
exponentially (full line), for finite $\zeta$ (dashed) saturates  to
the value $\propto 
\zeta$ of the energy difference between the minima. In this regime,
the tunneling of the oscillator between the minima as characterized by
the weight of the  
soft-mode peak in the phonon propagator (shown dotted) is suppressed.

\subsection{Conductance}
 \begin{figure}[t]%
\includegraphics*[width=0.8\linewidth,keepaspectratio]{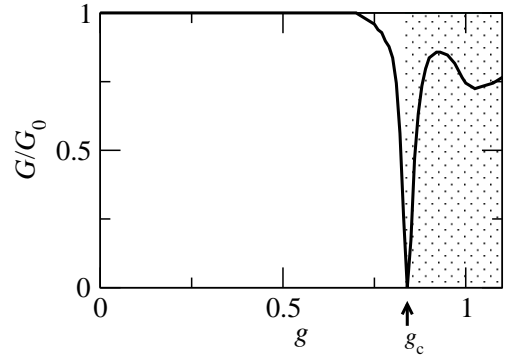}
\caption{%
Conductance, $\zeta=0.01$. Shaded area ($g>g_c$) indicates the unphysical
regime. 
\label{fig6}}
\end{figure}

We show the conductance calculated by NRG in Fig.~\ref{fig6}. In the
particle-hole symmetric point and at zero temperature the system is in the
unitary limit, hence the conductance (in units of $G_0$) is for $g=0$
near unity (reduced only due to small breaking of inversion symmetry
for a value of order $\zeta^2$). For increasing $g$ it decreases and
at $g \sim g_\mathrm{d}$ if drops to zero. This occurs because the
molecular orbital is increasingly hybridized only to the right lead
and the coupling to left lead $V(1-gx)$ becomes small. The point where
the conductance is zero corresponds to the decoupling of the left
lead, $V(1-gx) \sim 0$, on the average.

Surprisingly, for $g$
larger still, the conductance increases again.
This is due to the overlap with the
negative amplitude $V(1-gx)<0$ when states for which $1\pm gx < 0$ contribute
considerably to the transport. As discussed in more detail in
Ref.~\cite{mravlje08} this regime is an artifact of
linearization and cannot be observed in the
break-junction experiments. Here we remark only, that
in the large-$g$ regime the spin is screened by  
the odd channel. At the delimiting value of $g$ (which we denote by
$g_\mathrm{c}$) the spin is simultaneously
screened by the both channels and the resulting state is characterized
by the non-Fermi liquid two-channel Kondo fixed point in the NRG
flow. 

\section{Conclusions}
In this report we considered a metallic break-junction bridged by a
molecule. We were interested in the electron transport through the
break-junction, which is influenced by the modulation of tunneling
between the 
molecular orbital and the leads  due to the oscillations of the
molecule with respect to the leads. 

We
have shown that due to the electron-phonon coupling the harmonic
potential confining the molecule to the center-of-inversion evolves
to the double well effective potential. The change in the form of
confining potential is reflected in 
dynamical properties -- {\it e.g.} a low-frequency peak emerges in the
phonon propagator -- as well as in the static properties -- {\it e.g.}
the fluctuations of displacement are increased.

The emergence of the double
well effective potential makes the system strongly susceptible
to perturbations breaking the 
inversion symmetry. When the frequency of the soft-mode decreases
below the energy scale of such a perturbation, the molecule
is attracted to one of the leads.  As a consequence, the conductance
is suppressed. 

There are many points which we have not discussed in detail, including
the breakdown of the linearization and the physics of the two-channel
Kondo fixed point. The more comprehensive analysis of this interesting
system will appear elsewhere~\cite{mravlje08}.

\begin{acknowledgement}
We acknowledge the use of R. \v{Z}itko's NRG
http://nrgljubljana.ijs.si and T. Rejec's SG code. 
\end{acknowledgement}

%

%

\providecommand{\WileyBibTextsc}{}
\let\textsc\WileyBibTextsc
\providecommand{\othercit}{}
\providecommand{\jr}[1]{#1}
\providecommand{\etal}{~et~al.}







\end{document}